\documentclass[preprint,12pt]{elsarticle}
\usepackage{amssymb}
\usepackage{amsmath}
\usepackage[separate-uncertainty=true]{siunitx}
\usepackage[colorlinks=true,urlcolor=blue]{hyperref} 
\usepackage{caption}
\usepackage{subcaption}
\usepackage{booktabs}
\usepackage{adjustbox}
\usepackage{tabularx}
\usepackage{placeins}
\usepackage{graphicx}
\DeclareSIUnit\year{yr}
\usepackage{xurl}

\usepackage{xcolor}
\newcommand{\sm}[1]{#1}

\usepackage{verbatim}

\journal{Smart Energy}

\begin{document}

\begin{frontmatter}


%%\tnotetext[label0]{Corresponding author}
\author{Simon Malacek\corref{cor1}\fnref{label1,label2}}
\ead{simon.malacek@tugraz.at}
\ead[url]{https:\\iee.tugraz.at}
%\fntext[label2]{where}
\cortext[cor1]{Coresponding author.}

\title{Generating Building-Level Heat Demand Time Series by Combining Occupancy Simulations and Thermal Modeling}

%\author[label1,label2]{Simon Malacek}
%\author[label1]{TBD}
\author[label3]{José Portela}
\author[label1]{Yannick Werner}
\author[label1,label2]{Sonja Wogrin}

\affiliation[label1]{organization={ Institute of Electricity Economics and Energy Innovation
at Graz University of Technology},
            addressline={Inffeldgasse 18}, 
            city={Graz},
            postcode={8010}, 
            state={Austria},
            country={}}

\affiliation[label2]{organization={Reserach Center ENERGETIC},
            addressline={Inffeldgasse 18}, 
            city={Graz},
            postcode={8010}, 
            state={Austria},
            country={}}

 \affiliation[label3]{organization={Instituto de Investigación Tecnológica (IIT), Universidad Pontificia Comillas},
             addressline={Calle Alberto Aguilera 23},
             city={Madrid},
             postcode={28015},
             state={Spain},
             country={}}

%\author{Simon Malacek and TBD and Sonja Wogrin} %% Author name

%\affiliation{organization={Graz University of Technology},%Department and Organization
 %           addressline={Inffeldgasse 18}, 
  %          city={Graz},
   %         postcode={8010}, 
    %        state={Austria},
     %       country={}}

%% Abstract
\begin{abstract}
%===== rewritten version of the abstract =====
Despite various efforts, decarbonizing the heating sector remains a significant challenge. To tackle it by smart planning, the availability of highly resolved heating demand data is key.
Several existing models provide heating demand only for specific applications. Typically, they either offer time series for a larger area or annual demand data on a building level, but not both simultaneously. Additionally, the diversity in heating demand across different buildings is often not considered. To address these limitations, this paper presents a novel method for generating temporally resolved heat demand time series at the building level using publicly available data. The approach integrates a thermal building model with stochastic occupancy simulations that account for variability in user behavior. 
As a result, the tool serves as a cost-effective resource for cross-sectoral energy system planning and policy development, particularly with a focus on the heating sector. The obtained data can be used to assess the impact of renovation and retrofitting strategies, or to analyze district heating expansion.
To illustrate the potential applications of this approach, we conducted a case study in Puertollano (Spain), where we prepared a dataset of heating demand with hourly resolution for each of 9,298 residential buildings. This data was then used to compare two different pathways for the thermal renovation of these buildings.
By relying on publicly available data, this method can be adapted and applied to various European regions, offering broad usability in energy system optimization and analysis of decarbonization strategies.

\end{abstract}

\begin{graphicalabstract}
%Mandatory Graphical Abstract
%A graphical abstract is mandatory for this journal. It should summarize the contents of the article in a concise, pictorial form designed to capture the attention of a wide readership online. Authors must provide images that clearly represent the work described in the article. Graphical abstracts should be submitted as a separate file in the online submission system. Image size: please provide an image with a minimum of 531 × 1328 pixels (h × w) or proportionally more. The image should be readable at a size of 5 × 13 cm using a regular screen resolution of 96 dpi. Preferred file types: TIFF, EPS, PDF or MS Office files. You can view Example Graphical Abstracts on our 

\includegraphics{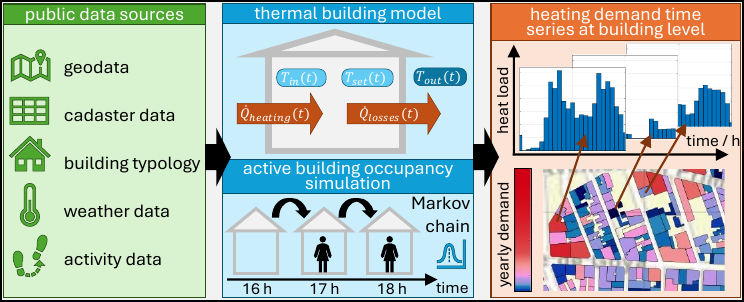}
%\textcolor{red}{To be done. Idea: Process sketch + map with yearly heat demand + inserts with time series of single buildings/apartments.} 
\end{graphicalabstract}

%Highlights are mandatory for this journal as they help increase the discoverability of your article via search engines. They consist of a short collection of bullet points that capture the novel results of your research as well as new methods that were used during the study (if any). Please have a look at the example Highlights.
%Highlights should be submitted in a separate editable file in the online submission system. Please use 'Highlights' in the file name and include 3 to 5 bullet points (maximum 85 characters, including spaces, per bullet point).

\begin{highlights}
\item This method provides high spatial and temporal resolution for heating demand time series.
\item Publicly available data creates real-world load profiles that consider diversity in heating demand among buildings.
\item Generated heating demand data supports energy planning, modeling, and analysis.
\end{highlights}

%% Keywords

\end{frontmatter}




\begin{table}[ht]
    \begin{tabularx}{\textwidth}{ll}
    \toprule
     \textbf{Abbreviations}    &  \\ \midrule
    GIS & geographic information system \\
    HDD & heating degree days \\
    INSPIRE & infrastructure for spatial information in the \\
    & European community \\
    LIDAR & light detection and ranging \\
    MCMC & markov-chain monte carlo \\
    \sm{NUTS} &\sm{nomenclature of territorial units for statistics} \\
    OSM & open street maps \\
    RC model & resistor-capacitor model \\
    TUS & time use survey \\   \bottomrule
    \end{tabularx}
    \label{tab:abbreviations}
\end{table}

\begin{table}[h]
    \begin{tabularx}{\textwidth}{ll}
    \toprule
     \textbf{Nomenclature}    &  \\ \midrule
     $i$    & index for buildings \\
     $t$   & index for the time steps \\
$A(t)$ & active occupancy at time step $t$ \\
\sm{$\mathrm{AHD}$} & \sm{annual heating demand} \\ 
$G$ & generalized thermal conductance \\
$k$ & generalized thermal storage capacity \\
$Q_\mathrm{demand}(t)$ & required heating demand for time step $t$ \\
$Q_\mathrm{heating}(t)$ & actual applied heating power for time step $t$ \\
$Q_\mathrm{heating,max}$ & maximum heating power \\
$Q_\mathrm{losses}(t)$ & thermal losses at time step $t$ \\
$Q_i$ & yearly heating demand for building i \\
$Q_\mathrm{spec,i}$ & specific heating demand for building i \\
$S_i$ & residential area of building $i$ \\
$T_\mathrm{in}(t)$ & indoor temperature for time step $t$ \\
$T_\mathrm{out}(t)$ & outside temperature for time step $t$ \\
$T_\mathrm{set,daytime}$ & setpoint temperature at daytime \\
$T_\mathrm{set,nighttime}$ & setpoint temperature at nighttime \\ \bottomrule
    \end{tabularx}
    \label{tab:symbols}
\end{table}

\FloatBarrier

\section{Introduction}
%The need for decarbonisation in the heat sector
Within the European Union (EU), the residential sector accounts for approximately \SI{28}{\percent} of the final energy consumption as of 2021~\cite{Eurostat2022_EnergyBalance}. Of this, more than \SI{78}{\percent} is used for space heating and hot water treatment~\cite{Eurostat2022_FinEnergyConHouseH}. Despite efforts to decarbonize the heat supply, \SI{77}{\percent} of this energy demand still comes from non-renewable sources~\cite{Eurostat2022_RESShare}, which contributes to about one-third of energy-related greenhouse gas emissions. The EU Energy Performance of Buildings Directive EU/2024/1275~\cite{energy_eff_directive} aims to reduce emissions from this sector by at least \SI{60}{\percent} by 2030 compared to 2015 levels, making it a crucial part of the EU's overall climate targets.

To address this challenge, the main pathways for reducing residential energy consumption are clear: 
\begin{itemize}
    \item improving building insulation to reduce heating demand~\cite{Walker2022},
    \item expanding district heating systems,
    \item and replacing oil- and gas-fired boilers with heat pumps alongside with increasing the share of renewables in the power system~\cite{Pastore2023}. 
    \end{itemize}
The required technologies and devices are available at a mature level and low costs~\cite{Jones2021}.

%show the problem
While the general strategy described above seems conceptually simple, its real-world implementation presents significant challenges. Boosting the thermal renovation of buildings and retrofitting heating systems requires well-targeted policies~\cite{DiazLopez}. Moreover, ongoing electrification of heating systems will increase the overall power demand, potentially leading to higher peak \sm{loads~\cite{Vaishnav2020}}. This could necessitate grid reinforcements to ensure stability and additional expansion of renewable power \sm{production~\cite{Ssembatya2024}.}

%show what ist needed to solve the problem 
Successfully transforming the building sector at minimal economic cost requires a smart, integrated planning and design approach~\cite{Rosenow2023}. This approach must consider the coupling of the heating and power sectors, as well as retrofitting measures that impact the distribution of loads. One critical requirement of this planning is access to accurate, high-resolution data on heating demand in both space and time domain~\cite{Lombardi2019}.

% show the advantages of having spatially and temporally resolved time series
On the one hand, high spatial resolution of heating demand data, e.g. at the building level, allows for the identification of the most effective renovation measures and enables the design of well-targeted policies like subsidy programs. This high spatial resolution is also essential when assessing the economic viability of district heating networks~\cite{Fritz2015}. 
On the other hand, high temporal resolution, e.g. hourly resolved heat demand time series, enables the analysis of the impact of heating system electrification on the power grid~\cite{Akmal2014}. By considering the additional electrical loads (e.g. from heat pumps) and analyzing the impact of thermal load shifting on the power system, these data can inform decisions for sector-coupled generation and expansion planning \sm{in integrated power and heat systems. For example, focusing on the integration of district heating systems \cite{schwappeSCGEPTEP} or energy hubs \cite{pypsa_case}.}
Together that shows that cost-efficient system planning and optimization strongly rely on granular data. This is often available (for research purposes) for the power sector as smart meter measurements at the building or household level, distribution network data, and national energy statistics, all with hourly or even quarter-hourly resolution. 
However, this level of detail is not yet available for heat load curves at any scale. Existing methodologies (discussed in detail in Section~\ref{sec:lit_review}) and data sources for heating demand provide either geolocated annual heating demand, synthetic load profiles for individual buildings, or aggregated load curves for specific regions based on gas or district heating consumption.

% showing the gap 
Only a few methodologies~\cite{Buettner2022,Heidenthaler2023,Malla2021} generate both hourly load profiles at a high spatial resolution. However, none of these approaches fully accounts for the stochastic nature of load profiles caused by diverse and asynchronous user behavior across different buildings. Furthermore, only a few are based solely on easily accessible open-source data. We discuss the most relevant methods in the following section.

\subsection{Review of existing methods in literature}
\label{sec:lit_review}
A wide range of approaches for modeling thermal demand can be found in the literature, each with different objectives. We summarize the key methods, focusing on spatially-oriented approaches in Section~\ref{sc:methods_spatial} and temporally-oriented methods in Section~\ref{sc:Intro-tempmehtods}. Finally, Section~\ref{sc:combindedapproches} explores approaches integrating spatial and temporal perspectives.

\subsubsection{Methods for spatial heating demand data}
\label{sc:methods_spatial}
%review of the current state-of-the-art models for heating demand data generation
%intro
The literature typically distinguishes top-down and bottom-up approaches for generating heating demand data~\cite{Frayssinet2018, Swan2009}. Top-down approaches, such as the one used by Gils et al.~\cite{Gils2013}, disaggregate the total heating demand of a defined region by weighting it with predefined parameters, most commonly population density. The total heating demand is either measured directly (e.g., through gas or electricity consumption) or derived from economic energy statistics. These models are relatively simple to implement when the necessary data is available. However, their spatial resolution is limited because statistical data cannot be traced down to the level of individual buildings, making them suitable for large-scale analyses but less effective for district-level planning tasks.

%bottom up approaches
With the increasing availability of open data sources, more bottom-up approaches have emerged. These models start at the smallest unit, usually buildings or even individual households. Geographic Information Systems (GIS) data, such as OpenStreetMap (OSM)~\cite{OSM}, are often utilized to geolocate buildings. Some studies also incorporate 3D building data (e.g., Schwanbeck et al.~\cite{Schwanebeck2021} for Germany) or LIDAR data (e.g.~\cite{Lumbreras2022} in Vitoria-Gasteiz, Spain) to perform more sophisticated analyses of building volumes. On the building level, the yearly heating demand is typically calculated based on residential area and a typical specific heating energy demand for that building type. Residential areas can be estimated from GIS data~\cite{Nielsen2013, Fallahnejad2018, Meha2021} or from cadastral and census information~\cite{MartinConsuegra2018}.

%required imput parameter for the bottum-up models
To obtain the specific heating energy demand for buildings, they are often categorized into different archetypes (i.e. types of representative buildings for a region). The TABULA Web Tool~\cite{Loga2016} is commonly used as the data source for this. The tool provides specific heating energy demand values (in \si{\kilo \watt \hour \per \year}) based on factors such as the country (which dictates national building standards), climate zone, building type (e.g., single-family house, terraced house, apartment block), and construction year. Some studies, like that of Dall'O' et al.~\cite{Dallo2012}, validate the building data through energy audits of sample buildings.

Beyond this standard approach, some models consider additional input parameters. For example, the Building Thermal Energy Assessment Model by Prades-Gil et al.~\cite{PradesGil2023} includes solar irradiation, which has been found to significantly impact specific energy demand, second only to climate variables~\cite{Staffell}. Martín-Consuegra~\cite{MartinConsuegra2018} also incorporates the facade area, allowing for higher spatial resolution even within building blocks.

%top-down and bottwom-up approaches
Several studies compare the results of bottom-up and top-down methods. Meha et al.~\cite{Meha2020} performed such a comparison for a test region in Kosovo. Calderón et al.~\cite{Calderon2015} calculated the end-use energy demand at the building level for the UK by combining top-down and bottom-up methods, focusing on national energy planning. Parmpuri et al.~\cite{Pampuri2017} employed a methodology that uses energy performance certificates from Switzerland’s database along with building age to calculate yearly demand. They validated the results by comparing the summed demands with statistical data on energy consumption at the canton level. All of these studies demonstrated a strong alignment between the top-down and bottom-up methodologies.

%available data bases / and maps
In addition to presenting different modeling approaches and individual case studies, several projects provide heating demand data at the European level. Notable examples include the Pan-European Thermal Atlas (PETA)~\cite{Moeller2021}, developed as part of the \textit{Heat Roadmap Europe} project~\cite{Persson2014}, and the sEEnergies project. The \textit{Hotmaps Toolbox}~\cite{Hotmaps} offers similar outputs and validates derived heating demands within reference cities. National projects, such as the \textit{MapaDeCalor}~\cite{MapDeCalor} for Spain, also provide useful heating demand data. While these maps allow for easy access to heating demands and related information, such as waste heat potentials at a regional level or integrated calculation tools, their spatial resolution is too coarse for district-level planning, where street or building-level resolution is required.

%different application of the spatial heating demand
The bottom-up approaches discussed above are particularly useful when actual measured heating demand data is unavailable. They are often applied to assess the potential for retrofitting~\cite{Meha2021} or to expand district heating systems. For instance, Nielsen and Möller~\cite{Nielsen2013} used a heat map (energy demand per building) of Denmark to assess the economic feasibility of district heating networks in new neighborhoods. Additionally, these approaches can be used to analyze energy poverty, as demonstrated by Terés-Zubiaga et al.~\cite{TeresZubiaga2023} at the building level. By focusing on retrofitting efforts, such studies can not only help achieve ecological goals but also generate social benefits by addressing energy poverty, reducing carbon emissions, and improving living conditions.

%review of modelling temporal heating demands
\subsubsection{Methods for temporal heat load profiles}
\label{sc:Intro-tempmehtods}
%overview
Various approaches exist for generating load profiles, which can be broadly categorized into data-driven black-box models and engineering white-box models based on physical equations and detailed individual building data. Further categorization of these approaches is discussed in the Supplementary Material of Staffell et al.~\cite{Staffell}. Simple, previously used, standard load profiles are no longer applicable for smart energy planning at the district level, as discussed in~\cite{Buettner2022}. An extensive review of the various methods, available software packages, and their different fields of application is provided by Jebaraj and Iniyan~\cite{Jebaraj2006} and more recently by Allegrini et al.~\cite{Allegrini2015}. Peacock's review from 2021~\cite{Peacock2021} analyses four common modeling approaches, including the Heating Degree Days (HDD) method, and compares them with actual measured data. The review shows that all methods yield good results, but they only provide daily temporal resolution.

%data analysis / data driven models
Data-driven models require minimal information about the actual physical parameters of buildings. However, they do rely on measured heating demands, typically obtained from existing district heating networks or gas consumption data~\cite{Koene2023}. These models often aim to predict future consumption in existing district heating networks or perform correlation analyses between heating demand and external variables such as temperature. For example, \sm{Dang et al.} analyzed heat load patterns in existing district heating systems, emphasizing the importance of considering not just thermal building parameters and outside temperature \sm{~\cite{Dang2022}}, but also, \sm{in a subsequent work}, user behavior for accurate predictions \sm{\cite{Dang2023}}. Similarly, Maljkovic and Basic~\cite{Maljkovic2020} explored the influence of various parameters on heating demand using machine learning, measured data, and surveys. Fumo and Biswas~\cite{Fumo2015} provide a review of regression analysis for predicting residential energy consumption, highlighting the simplicity of regression models compared to physical models for demand prediction. However, when no data is available, one must rely on engineering or physical models instead.

%physicsl models
Engineering models, such as EnergyPlus~\cite{Crawley2001}, are well-established tools for detailed simulations based on energy balance equations and physical parameters. These models can provide sub-hourly resolved load profiles for individual buildings. However, due to the need for highly detailed parameters, these models are not suitable for whole cities or large areas. Additionally, they typically do not account for varying occupancy patterns across different buildings. As a result, simply adding up the simulated profiles of multiple buildings (as in~\cite{Heidenthaler2023}) can lead to unrealistic aggregated load peaks. 

%stochastic models
To address the issue of artificial load peaks, stochastic models have been introduced. Palacios-Garcia et al.~\cite{PalaciosGarcia2018} present a stochastic modeling and simulation approach for (electric) heating and cooling demand. Fischer et al.~\cite{Fischer2016} also use a stochastic bottom-up model to estimate load profiles for space heating and hot water demands. Stochasticity in these models refers to incorporating randomness alongside physical modeling, which better reflects the individual asynchronous behavior of building occupants.

%the value of considering user behaviour
Stochastic properties can be implemented using the Markov-Chain Monte Carlo (MCMC) method, a standard approach for random sampling. One of the first to incorporate this into load profile synthesis was Richardson et al.~\cite{Richardson2008}, who emphasized the importance of user behavior in accurately modeling heating and electricity demand. They proposed a methodology based on Time Use Surveys (TUS)~\cite{TUS_data} to generate active occupancy profiles for integration into demand models. A similar approach was employed by Ding et al.~\cite{Ding2023} to generate temporal occupancy profiles even on a room level. Furthermore, a more sophisticated analysis of occupancy behavior, considering demographic and household characteristics, was discussed by Fu et al.~\cite{Fu2022}, suggesting a more detailed approach to modeling occupancy patterns.

%databeses
At the database level, the \textit{When2Heat} project \sm{(original publication~\cite{Ruhnau2019}, current version of the dataset~\cite{Ruhnau2023})} provides demand data and other parameters, such as the coefficient of performance for heat pumps, for several European countries. This comprehensive database offers quick and easy access to valid load profiles. However, these data are aggregated at a \SI{0.75}{\degree} x \SI{0.75}{\degree} grid level, which does not provide sub-district level spatial resolution.

\subsubsection{Combined approaches}
\label{sc:combindedapproches}
%recgonised the need for temporal and spatial resolved heating demands
Several studies~\cite{Berger2018,Koene2023,Buettner2022} have discussed the advantages of obtaining heating demand data with both high spatial and temporal resolution, presenting various approaches to achieve this.

Berger and Worlitschek~\cite{Berger2018} provide a case study with heating demand data at a spatial resolution of one square kilometer for all of Switzerland, but with a temporal resolution limited to one day, based on the  HDD method. They propose a top-down methodology to generate aggregated load curves, combining population density distribution maps, norm temperature profiles, HDDs, and total (measured or statistical) residential heating demand. In contrast, Koene and Eslami-Mossallam~\cite{Koene2023} use an electrical equivalent resistor-capacitor (RC) model to generate hourly heating load profiles, comparing them with gas consumption data. However, the spatial resolution is limited to the district level, and thermal properties are estimated from the gas demand signature (top-down), rather than being derived from individual building data. This approach is therefore unsuitable for districts with no known gas or district heating demand, such as those without district heating systems or those utilizing diverse decentralized technologies like wood, oil ovens, and heat pumps.

Some studies offer spatially and temporally resolved data for specific regions, but not at the building level: Malla~\cite{Malla2021}, in his master's thesis, recognizes the need for high spatial and temporal resolution in heat demand profiles and presents a methodology to calculate load profiles for a \SI{100}{\metre} x \SI{100}{\metre} grid across Germany. Clegg and Mancarella~\cite{Clegg2019} developed a method with a half-hourly resolution for 404 areas of Great Britain. While they validate their time series against measured gas consumption data, the spatial resolution remains at the district level. Lombardi et al.~\cite{Lombardi2019} present a thermodynamic modeling approach for Italian regions, incorporating many physical details for heat transfer to build an accurate model. However, the spatial resolution is limited to the NUTS-2 \cite{nuts} level \sm{(corresponds to regions/provinces)}. Their model (with available code) also enables running different refurbishment scenarios defined by the user.

Recently, Staffell et al. proposed a global model for hourly heating and cooling demands~\cite{Staffell}, which provides temporally resolved load profiles for locations worldwide. However, not on a building-level. These profiles are publicly available on~\cite{RENW_Ninja}, though they need to be scaled according to the actual yearly heating demand of the region of interest. 

An interesting approach by Büttner et al.~\cite{Buettner2022} applies a top-down methodology, disaggregating census data down to the building level. Each building is assigned a pre-generated load profile from a set of 1,259 different profiles based on its properties. However, this top-down approach distributes heating demand evenly across all buildings within a given cell, regardless of the actual building type. As a result, the method is not capable of capturing variability between buildings within a district.

Finally, Heidenthaler et al.~\cite{Heidenthaler2023} uses energy performance certificates and a simulation tool to generate load profiles at the building level. However, this model does not account for the stochastic nature of user behavior, leading to unrealistic peaks in heating demand.

\subsection{Original contribution}
%original contribution 

\sm{The literature review highlights the variety of existing methods and approaches, each with a specific scope of application and inherent limitations. For instance, traditional building energy models can produce highly precise and, when properly calibrated, accurate heat demand time series for individual buildings or small building clusters. However, they require detailed input data -- such as building geometry, orientation, wall structure, construction materials, and insulation characteristics -- which may not be readily available for large-scale applications. In contrast, regression-based models can provide reasonable estimates at aggregated levels (e.g., district or city scale), but they typically lack the spatial granularity required to resolve heat demand at the individual building level.}

\sm{To the best of our knowledge, none of the existing approaches can generate stochastic heat demand time series at the building level using only publicly available data for a large number of buildings. While individual features have been implemented into existing models, none of them combine all these elements into a unified framework.  However, such integration is essential to enable the straightforward generation of heat demand data in data-scarce contexts, supporting a wide range of applications in energy system modeling -- particularly those focusing on sector coupling with high spatial resolution, such as the planning of local heat networks for waste heat utilization. To address this methodological gap, we present a novel approach that integrates the following distinct features into a single workflow:}

\begin{enumerate}
    \item \sm{\textbf{Publicly Available Data Utilization} – Tailoring the workflow to work exclusively with publicly available data ensures broad applicability and superior scalability, enabling analysis of thousands of buildings and overcoming data shortages.}
    \item \sm{\textbf{Simplified Physics-Based Thermal Model} – Using a physics-based model allows for the generation of reasonable demand profiles at the individual building level, making the method independent of standard load profiles, disaggregation techniques, or previously measured profiles.}
    \item \sm{\textbf{Stochastic Occupancy Simulation} – Incorporating user behavior and utilizing stochastic simulation ensures realistic demand profiles that account for household asynchronicity and demand fluctuations, while maintaining scalability from a single neighborhood to entire regions.}
\end{enumerate}

\sm{This novel approach enables the flexible generation of high-resolution heat demand data with minimal input requirements -- supporting the decarbonization of heat supply across diverse regions. Although showcased through a specific case study, the method is widely applicable and highly scalable, offering a powerful tool for data-driven energy system planning.}

\subsection{Organization of the remainder of this paper}
 %short outlook about the main contents of the paper
Section~\ref{sc:methodology} provides a detailed explanation of the data sources and methods used, followed by validation results in Section~\ref{sec:data-validation}. The subsequent case study (Section~\ref{sc:casestudy}) applies the methodology to a city in Spain, highlighting the importance of the resulting data and its potential applications. The discussion on these applications, along with the limitations and possible areas for further improvements, can be found in Section~\ref{sc:conclusion}. 

\section{Methodology}
\label{sc:methodology}
In this section, we present a methodology to generate hourly heat demand profiles on the single building level based on publicly accessible data and considering the stochasticity of user behavior. \sm{The process is designed to work with a minimum input of publicly available data, ensuring broad applicability. By that, it provides valuable data for various energy analyses across different fields of research, particularly where this data is currently lacking or unavailable.}

After discussing different data sources in~\ref{sc:datasources}, each step of the model is explained in detail in Subsections~\ref{sec:step1} -~\ref{sec:step6}. Subsequently, Subsection~\ref{sec:data-validation} illustrates the validity of the method using several comparative and analytic approaches.  

\subsection{Data sources}
\label{sc:datasources}
Table~\ref{tab:data-sources} provides an overview of the required data for the proposed model and lists publicly available data sources. The references cited in the \textbf{Source} column of Table~\ref{tab:data-sources} provide data within the EU. However, the accuracy and completeness of the data can vary for different countries and locations. 
Any other source providing similar data can be used for the following steps. 

\begin{table}[t]
    \centering
    \caption[Data sources]{Data sources for the subsequent generation of spatial-temporal resolved heat demand profiles.}
    \label{tab:data-sources}
    \begin{adjustbox}{max width=\textwidth}
    \begin{tabular}{lll}
        \textbf{Type} & \textbf{Description} & \textbf{Source} \\ \toprule
        geodata  & location and shape of buildings & Open Street Maps~\cite{OSM} \\ \midrule
        cadaster data & number of units and number of floors,& INSPIRE~\cite{geodata_directive} \\
        &year of construction, (residential) area, & \\ 
        &address, type of usage &  \\ \midrule
        building typology & specific heating demand & TABULA WebTool~\cite{Loga2016} \\ \midrule
        active occupancy & activity statistics and time profiles & TUS~\cite{TUS_data} \\ 
        profiles \\ \midrule
        weather data & hourly resolved outdoor temperature & Renewable.ninja~\cite{RENW_Ninja}  \\ \bottomrule        
    \end{tabular}
    \end{adjustbox}
\end{table}

\subsection{Data processing}
\begin{figure}
    \centering
    \includegraphics[width=\textwidth]{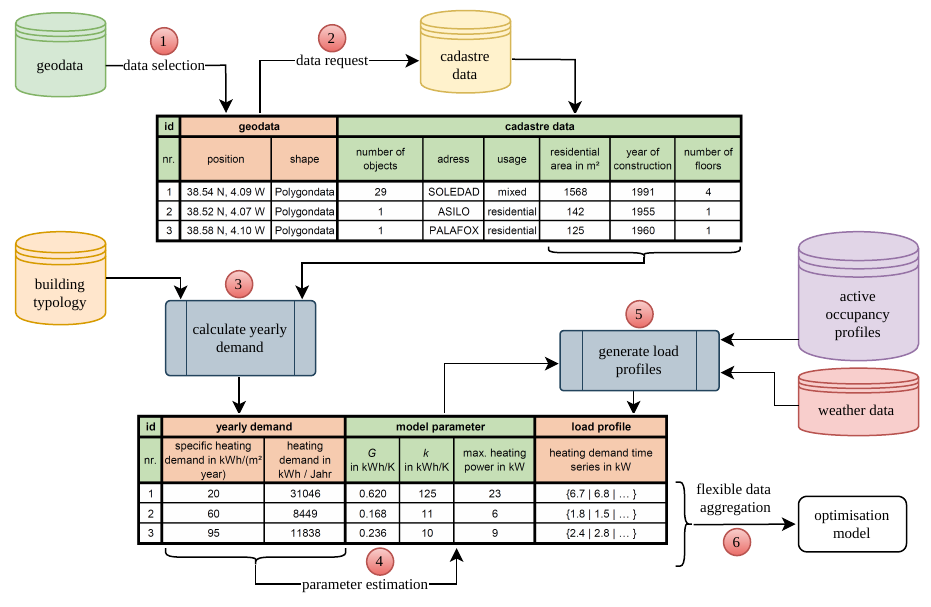}
    \caption[Illustration of the data processing workflow]{Schematic illustration of required data sources and the data processing workflow. The enumerated steps correspond to the steps in Section~\ref{sec:step1}.}
    \label{fig:workflow_dataprocessing}.
\end{figure}

The following paragraphs describe the data processing, which can be subdivided into the following six steps: 
\begin{enumerate}[Step 1: ]
    \item Acquire geo-data, including the buildings with shape information and GPS location.
    \item Request data from a cadaster/census data source for every building.
    \item Categorize every building according to a building typology.
    \item Estimate the thermal properties of every building. 
    \item Generate load profiles by considering user behavior by MCMC method and the thermal building model.
    \item Aggregate data for subsequent applications (if necessary).
\end{enumerate}
Figure~\ref{fig:workflow_dataprocessing} illustrates the corresponding workflow and data sources. 

% brief description of how to get the geodata, what is in it, and how to store them.
\subsubsection{Step 1: Geo-data conditioning}
\label{sec:step1}
To begin, the region of interest for the analysis is identified. OpenStreetMap (OSM)~\cite{OSM} serves as a robust source of geospatial data, providing structured geoinformation. This typically includes a (geo)object for each building, containing at least its centroid GPS location and shape, and often additional attributes such as postal address, building height, or usage type. Moreover, OSM frequently covers other infrastructure, such as streets and electricity grids, which can be valuable for further analysis. Data for the selected region can be downloaded directly or obtained using tools like Overpass Turbo~\cite{overpass_turbo}, which facilitates additional filtering and preprocessing. Subsequently, the structured geodata must be transformed into a tabular format, where each row corresponds to a building and includes its associated parameters.

\subsubsection{Step 2: Cadaster data request}
\label{sec:step2}
% discussion how data can be requested (INSPIRE) EU projects, and which data is needed. 
The EU directive 2007/2/EG, passed in 2007~\cite{geodata_directive}, aims to ensure the provision of comprehensive geospatial information across the EU. To achieve this, INSPIRE (Infrastructure for Spatial Information in the European Community) was established, enabling access to cadastral data in all member states. While the specific implementation varies by country and region, options such as bulk downloads or individual data requests based on address, cadastral reference, or GPS location are generally available. Ideally, these datasets provide information at the building level, including the number of units, usage type, year of construction, and (residential) area $S_i$. If such data is unavailable, it can be approximated using alternative parameters: \sm{The area can be estimated from the building’s shape and height, as demonstrated in Ref.~\cite{Lumbreras2022}. For buildings with missing construction year data, a default yearly heating demand, representing a weighted average, is assigned in the subsequent steps. These fallback assumptions enhance the method’s robustness against missing data entries. However, each additional estimation introduces a potential source of uncertainty. To quantify this error margin, the Supplemental Material~\cite{supply} compares results from cadastral data with the simplified estimation based solely on OSM, showing that the difference in yearly heating demand amounts to \SI{14}{\percent}.} 
To apply our method, this information is retrieved from the cadastral records and appended to each building entry in the dataset.

\subsubsection{Step 3: Determination of yearly heating demands}
\label{sec:step3}
% describe the building typology as in tabula 
A straightforward approach identifies three key parameters that primarily influence a building's heating energy demand: (i)~\textbf{Location:} The country and climate zone determine the climatic conditions, including the outside temperature and heating degree days. Additionally, the building's energy performance standards are often tied to its geographic location.
(ii)~\textbf{Year of Construction:} This reflects the thermal performance legislation, typical building standards, and construction methods in place at the time the building was erected.
(iii)~\textbf{Building Type:} The classification (e.g., single-family house, terraced house, apartment block) influences the volume-to-envelope ratio. A higher ratio, indicative of a more compact design, generally results in better thermal performance.
The TABULA Web Tool~\cite{Loga2016} supports this analysis by providing a detailed typology that categorizes buildings based on these parameters and offers specific heating energy demand estimates.

Using this data, each building in the dataset can be classified, yielding a specific heating demand, $Q_{\mathrm{spec},i}$, for each building. The total residential heating demand, $Q_i$, for a building is then calculated by scaling the specific heating demand with its residential area, $S_i$, using: $Q_i = S_i Q_{\mathrm{spec},i}$. 

\subsubsection{Step 4: Estimation of thermal properties}
\label{sec:step4}

\begin{figure}
    \centering
    \includegraphics[]{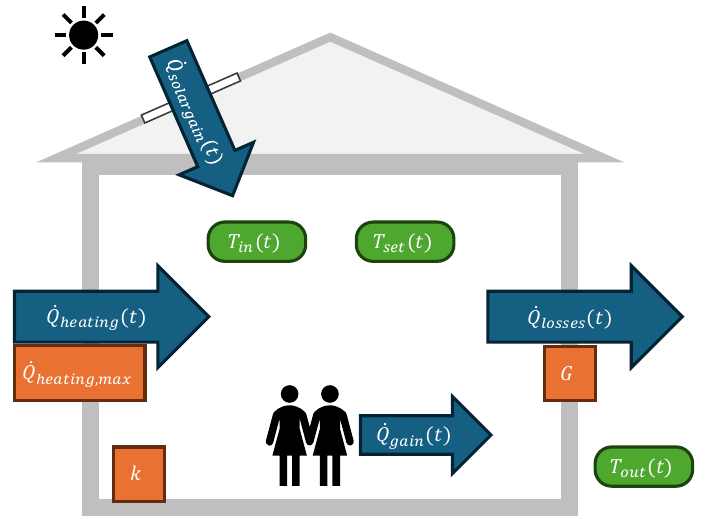}
    \caption[Thermal model for generating heating demand load profiles]{Thermal model for generating heating demand load profiles. $T_\mathrm{in}(t)$ = hourly actual indoor temperature \sm{in \si{\celsius}}, $T_\mathrm{out}(t)$ = hourly outdoor temperature \sm{in \si{\celsius}}, $T_\mathrm{set}(t)$ = hourly set temperature based on occupancy and daytime in \sm{\si{\celsius}}, $k$ = thermal \sm{storage capacity in \si{\kilo \watt \hour \per \kelvin}}, $G$ = thermal conductance \sm{in \si{\kilo \watt \per \kelvin}}, $\dot{Q}_\mathrm{heating,max}$ = maximum heating power \sm{in \si{\kilo \watt}}, $\dot{Q}_\mathrm{losses}(t)$ = heat losses \sm{in \si{\kilo \watt}}, $\dot{Q}_\mathrm{gain}(t)$ = heat gain by internal sources \sm{in \si{\kilo \watt}},  $\dot{Q}_\mathrm{solargain}(t)$ = heat gain by solar irradiation \sm{in \si{\kilo \watt}}, $\dot{Q}_\mathrm{heating}(t)$ = actual heating power \sm{in \si{\kilo \watt}}.}
    \label{fig:thermal_model}
\end{figure}

In real-world scenarios involving large numbers of buildings, detailed data on construction materials, wall thicknesses, and window types is often unavailable. To address this, we implemented a simplified thermal model, as illustrated in Figure~\ref{fig:thermal_model}.

In this model:
\begin{itemize}
    \item \textbf{Thermal losses}, including transmission through walls, floors, roofs, and ventilation, are \sm{represented by} a single parameter \sm{for each building $i$}: the thermal conductance $G_i$ (in \si{\kilo \watt \per \kelvin}). \sm{In the absence of more detailed data,} this \sm{parameter} is estimated by dividing the total annual heating demand \sm{for the respective building $\mathrm{AHD}_i$} by the cumulative indoor-outdoor temperature difference over all hours of the year\sm{: 
    \begin{equation}
         G_i = \frac{\mathrm{AHD_i}}{\sum\limits_{t \mid T_{\mathrm{set}}(t) > T_{\mathrm{out}}(t)} \left( T_{\mathrm{set}}(t) - T_{\mathrm{out}}(t) \right)} ~~ \forall i
    \end{equation}}

    \item \textbf{Thermal storage capacity}, $k$ (in \si{\kilo \watt \hour \per \kelvin}), is assumed based on average values for typical construction materials and wall thicknesses~\cite{Goeke2021}, scaled according to building size. While more detailed distinctions could be made using building type and construction year, the marginal improvement in accuracy may not justify the significantly increased effort to track specific materials for every building.
    \item \textbf{Maximum heating power}, $\dot{Q}_\mathrm{heating,max}$ (in \si{\kilo \watt}), is determined using a standard method for heating system dimensioning~\cite{Heizlastberechnung}, independent of the heating supply type.
\end{itemize}

The simplifications outlined for estimating thermal properties are intended to ensure broad applicability across diverse scenarios. However, if more detailed data is available for a specific analysis, it can be incorporated to replace the simplified assumptions and enhance the accuracy of the results.

\subsubsection{Step 5: Synthetic load profile generation}
\label{sec:step5}

The hourly heating demand is influenced not only by the thermal properties of the building and the outside temperature but also significantly by the indoor setpoint temperature. For this analysis, we assume the setpoint temperatures of individual dwellings follow a normal distribution around $T_\mathrm{set,daytime} = \SI{21}{\celsius}$ for daytime and $T_\mathrm{set,nighttime} = \SI{16}{\celsius}$ for nighttime with a standard deviation of \SI{\pm 2}{\celsius} for both. To account for the temporal variation in setpoint temperatures, we introduce the concept of active occupancy. When residents are at home and active, the temperature is set to $T_\mathrm{set,daytime}$; otherwise, it is reduced when residents are either away (e.g., at work or school) or inactive (e.g., sleeping).

This behavior is represented by a Boolean variable $A(t)$:
\begin{itemize}
    \item $A(t) = \text{true}$ indicates the residents are active and at home.
    \item $A(t) = \text{false}$ means the residents are either away or inactive.
\end{itemize}

The general distribution of active occupancy can be derived from Time Use Surveys~\cite{TUS_data}, which are regularly conducted across European countries. These surveys capture daily routines through empirical questionnaires administered to a statistically representative sample population.
However, applying a single, uniform time profile for all buildings in the thermal model would not accurately reflect real-world variability, as people do not wake up or follow their daily routines at the same time. To address this, the MCMC method~\cite{Richardson2008} is used to generate individual active occupancy profiles $A_i(t)$ for each dwelling $i$.
In an MCMC chain, the next state $A(t+1)$ depends only on the current state $A(t)$ and is determined by a transition matrix specific to the time step, with a randomness factor introduced for transitions. Different transition matrices are used for weekdays and weekends to reflect workday-dependent behavior. The randomness in the generated profiles captures individual variability in daily routines. Nevertheless, the average of a sufficiently large number of profiles converges to the distribution derived from TUS data. \sm{A detailed implementation of the MCMC method, including sample data and validation, can be found in the Supplementary Materials~\cite{supply}.}

Figure~\ref{fig:MCMC-process} illustrates this process, showing individual generated profiles (Subfigure~\ref{singleMCMCprof}) and the average of many profiles (Subfigure~\ref{aggregMCMCprof}), converging for large sample numbers to the overall desired distribution. \sm{It should be noted that due to the stochastic nature of MCMC, the individual profiles will differ with each simulation run. However, this does not impact the aggregated results. In cases where reproducibility is desired (e.g., to obtain consistent numbers for testing), each building can be assigned a seed number, to produce a reproducible series of random numbers.}

\begin{figure}
    \centering
    \begin{subfigure}[b]{0.45\textwidth}
        \centering
        \includegraphics[width=\textwidth]{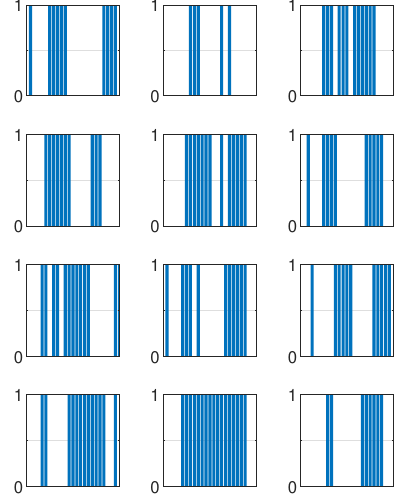}
        \caption{Occupancy profiles for 12 single buildings for one day (0 - 24 h on the x-axis), generated by MCMC.}
        \label{singleMCMCprof}
    \end{subfigure}
    \hfill
    \begin{subfigure}[b]{0.45\textwidth}
        \centering
        \includegraphics[width=\textwidth]{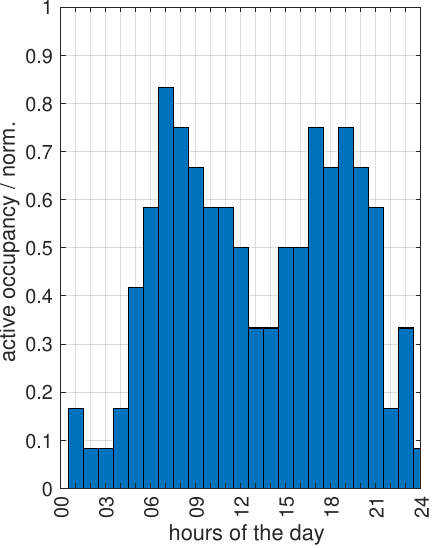}
        \caption{Average active occupancy by aggregation of the 12 single building profiles from the left figure.}
        \label{aggregMCMCprof}
    \end{subfigure}    
   
    \caption{Illustration of the convergence of single discrete active occupancy profiles (a), towards an average active occupancy profile (b).}
    \label{fig:MCMC-process}
\end{figure}

Once the active occupancy profiles are determined, the temperatures and thermal flows can be calculated using a quasi-static, time-discrete model with hourly time steps $\Delta t$. The thermal losses $\dot{Q}_\mathrm{losses}(t)$ for each hour are computed using Eq.~\eqref{eq:losses}, which is based on the difference between the current indoor temperature and the outdoor temperature, multiplied by the generalized thermal conductance $G$:

\begin{equation}
    \dot{Q}_\mathrm{losses}(t) = G\cdot(T_\mathrm{in}(t) - T_\mathrm{out}(t))
    \label{eq:losses}
\end{equation}

Having determined the active occupancy profiles, the setpoint temperature for each time interval $T_\mathrm{set}$ is directly derived from the active occupancy and the statistically distributed parameters $T_\mathrm{set,daytime}$ and $T_\mathrm{set,nighttime}$, as shown in Eq.~\eqref{eq:settemp}, which are fixed for each model run.

\begin{equation}
    T_\mathrm{set} = \begin{cases}
			T_\mathrm{set,daytime}, & \text{if } A(t) = \text{true}\\
            T_\mathrm{set,nighttime}, & \text{if } A(t) = \text{false}
		 \end{cases}
   \label{eq:settemp}
\end{equation}

To calculate the actual current indoor temperature $T_\mathrm{in}(t)$, the total heat provided by heating $\dot{Q}_\mathrm{heating}(t-1)$, plus internal gains $\dot{Q}_\mathrm{gain}(t-1)$ \sm{and solar gains $\dot{Q}_\mathrm{solargain}(t-1)$}, minus the thermal losses $\dot{Q}_\mathrm{losses}(t-1)$ from the previous time step, is divided by the thermal capacity $k$ of the building and added to the previous indoor temperature $T_\mathrm{in}(t-1)$ (see Eq.~\eqref{eq:Tin-iterative}). Internal gains arise from \sm{the use of} electrical appliances, cooking, \sm{or the body heat of occupants}. \sm{Depending on the available information and the required level of detail, a constant contribution can be assumed \cite{Koene2023}, the profile can be linked to active occupancy \cite{Fischer2016}, or it can be derived from empirical measurements \cite{Firlag2013}. In our approach, the magnitude of internal gains per building is scaled by the number of dwellings.} \sm{Solar gains account for the thermal energy received by the building from solar radiation. This contribution can be roughly estimated using hourly irradiation and window area, as discussed in \cite{Frayssinet2018} and Section \ref{sec:uncert}.} Initial values of $\dot{Q}_\mathrm{heating}(t=0) = 0$ and $T_\mathrm{in}(t=0) = T_\mathrm{set}$ are chosen.

\begin{equation}
    T_\mathrm{in}(t) = T_\mathrm{in}(t-1) + \Delta t ~ \frac{\dot{Q}_\mathrm{heating}(t-1) + \dot{Q}_\mathrm{gain}(t-1) \sm{ +\dot{Q}_\mathrm{solargain}(t-1)} - \dot{Q}_\mathrm{losses}(t-1)}{k}
    \label{eq:Tin-iterative}
\end{equation}

Knowing the indoor temperature, the required amount of energy $\Delta Q(t)$ to heat the building to $T_\mathrm{set}$ can be calculated as the thermal capacity $k$ times the temperature difference (see Eq.~\eqref{eq:DelatQ}).

\begin{equation}
    \Delta Q(t) = k \cdot (T_\mathrm{set}(t) - T_\mathrm{in}(t))
    \label{eq:DelatQ}
\end{equation}

By adding the thermal losses, the total heating demand for that particular time step $\dot{Q}_\mathrm{demand}(t)$ is calculated as:

\begin{equation}
    \dot{Q}_\mathrm{demand}(t) = \dot{Q}_\mathrm{losses}(t) + \dfrac{\Delta Q(t)}{\Delta t} 
    \label{eq:demand}
\end{equation}

Finally, the actual heating demand is determined by a case differentiation in Eq.~\eqref{eq:heating-cases}. \sm{If the calculation results in a negative value for a particular hour, it is set to zero. Otherwise, a negative heating demand would imply a decrease in indoor temperature, which does not reflect real-world conditions.} Additionally, a maximum heating power $\dot{Q}_\mathrm{heating,max}$ is introduced to limit the added heat per unit time, better representing a real system setup.

\begin{equation}
    \dot{Q}_\mathrm{heating}(t) = \begin{cases}
        0 & \text{if } \dot{Q}_\mathrm{demand} < 0 \\
        \dot{Q}_\mathrm{demand}(t) & \text{if } 0 < \dot{Q}_\mathrm{demand} < \dot{Q}_\mathrm{heating,max}  \\
        \dot{Q}_\mathrm{heating,max} & \text{if } \dot{Q}_\mathrm{demand} > \dot{Q}_\mathrm{heating,max} 
    \end{cases}\
    \label{eq:heating-cases}
\end{equation}

\begin{figure}
    \centering
    \includegraphics[width=0.95\linewidth]{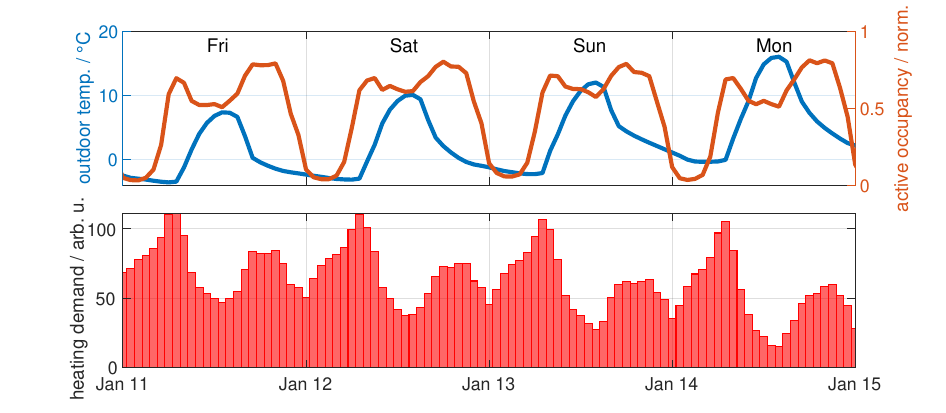}
    \caption{Relation between heating demand and the outside temperature and active occupancy. Peaks in the heating demand correlate with low outside temperature and high active occupancy, leading to the characteristic morning and evening peaks.}
    \label{fig:TempOccupHD_Relation}
\end{figure}

Through iterative calculations, a time series (e.g., for one year with 8760 hours) can be generated, yielding an individual heating demand $\dot{Q}_\mathrm{heating}$ for every building. Figure~\ref{fig:TempOccupHD_Relation} illustrates an example of such a heating demand profile and its dependency on the outside temperature and active occupancy.

\subsubsection{Step 6: Flexible data aggregation}
\label{sec:step6}
The heating demand profiles obtained at the building level can now be directly applied for further data analysis, or incorporated into simulations or optimization models. Using one of the various methods for network generation, such as minimum spanning tree \sm{for previously clustered buildings~\cite{Unternaehrer2017} or at the individual building level~\cite{Lumbreras2022}}, a district heating network topology can be proposed. Based on this topology, the load profiles can be easily aggregated by summing all the profiles from a particular branch, thereby drastically reducing computational effort.

Additionally, these thermal load profiles can be translated into electricity demands by assuming the use of electric boilers or heat pumps for heating in the building. The high spatial resolution of the data allows for the assignment of the additional power demand to specific distribution grid nodes. This enables an analysis of the impact of expansion decisions for the heating system (such as whether to connect to the district heating network or use a heat pump or electric boiler) on the overall system. Both the electric grid and the district heating system, along with their respective thermal demands, are accounted for. This approach facilitates the identification of the cross-sectoral optimal solution for each building.

\sm{While the annual heating demand calculated by our method is deterministic for each individual building, the corresponding time series exhibit stochastic variations due to the MCMC-based occupancy modeling. By aggregating time series, individual patterns are smoothed out, leading to a convergence towards a common distribution. Since only the active occupancy profiles themselves have a stochastic component, it can be assumed that the stochastically influenced part of the heat profiles will converge in the same manner (i.e., by aggregating approximately 10 dwelling profiles, as shown in the Supplementary Material~\cite{supply}). Any remaining differences in the aggregated heat demand time series are primarily caused by variations in the respective building stock.}

\subsection{Validation}
\label{sec:data-validation}

We validate the results obtained by our method in three key ways: First, by comparing the results with those of another study in~\ref{sec:replicationstuy}, we demonstrate the reasonableness of the heating demand at the building level. Second, in~\ref{sec:correlanalysis}, we show the correlation between heating demand and temperature, confirming the correct distribution of the daily heating demand throughout the year. Finally, in Section~\ref{sec:dailyprofiles}, we compare the hourly intra-day profiles generated by our method to demonstrate their strong agreement with the intra-day distributions reported in the literature.

\subsubsection{Replication analysis}
\label{sec:replicationstuy}
\begin{figure}
    \centering
    \includegraphics[width=\textwidth]{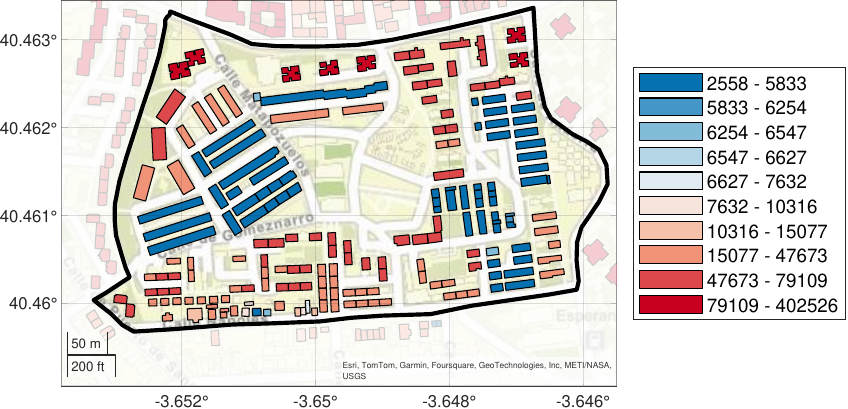}
    \caption{Yearly heating demand on building level in \si{\kilo \watt \hour \per \year}. For better comparison, the color scale was adapted to Ref.~\cite{MartinConsuegra2018}.}
    \label{fig:replication-study}
\end{figure}

To illustrate that our methodology retrieves the correct yearly heating demand at the building level, we applied it (up to step 3, see Section~\ref{sec:step3}) to a quarter in Madrid, which was analyzed in a study by Martín-Consuegra (Ref.~\cite{MartinConsuegra2018}). Figure~\ref{fig:replication-study} shows our results. For better comparison, the color codes were selected according to Figure 6 in the original publication Ref.~\cite{MartinConsuegra2018}. Although fewer details at the sub-building level are resolved, the heating demand at the whole building level aligns well, demonstrating good accordance between the methodologies at the single-building resolution. 

\subsubsection{Correlation analysis}
\label{sec:correlanalysis}
\begin{figure}
    \centering
    \includegraphics[width=\textwidth]{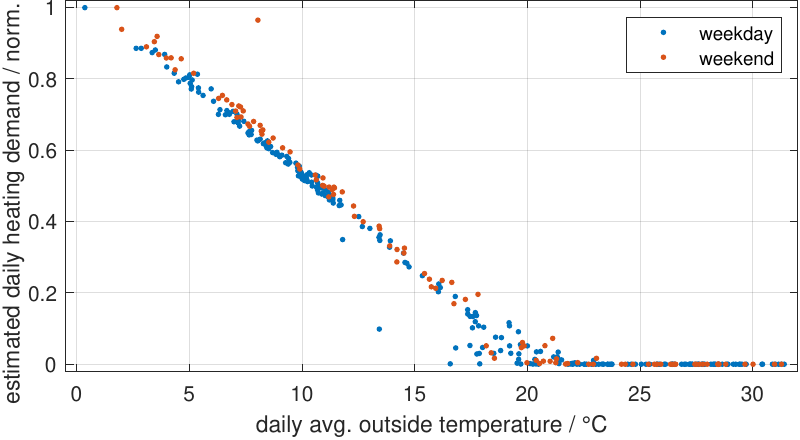}
    \caption{Correlation analysis between estimated daily heating demand and daily average outside temperature.}
    \label{fig:correlation-analysis}
\end{figure}

The daily heating demand shows a strong correlation with the daily average outside temperature, as described in the literature, such as in Ref.~\cite{Staffell}. Figure~\ref{fig:correlation-analysis} demonstrates that this expected correlation can also be found in the data obtained through our methodology. Additionally, accounting for occupancy results in higher heating demand on weekends compared to weekdays with similar outside temperatures. Conversely, the variability in heating demand for days with similar outside temperatures is attributed to the thermal inertia of the buildings, which is reproduced by our model. Together, these findings suggest reasonable results at a daily-based temporal scale.

\subsubsection{Comparison of daily profiles}
\label{sec:dailyprofiles}
\begin{figure}
    \centering
    \includegraphics[width=1\linewidth]{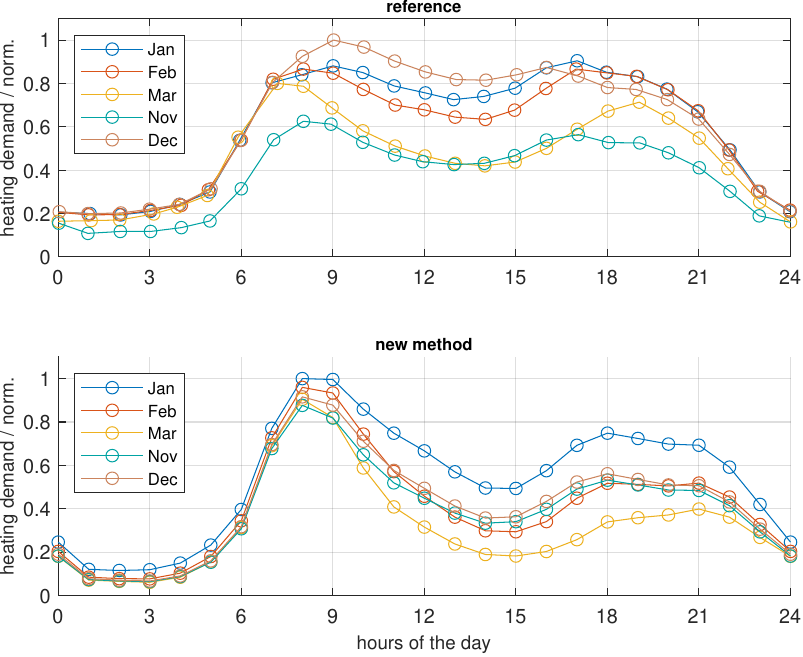}
    \caption{Comparison of average daily heating demand profiles. The upper tile (reference), shows actual measured heating demand data from Ref.~\cite{Koene2023}, in comparison, the lower tile shows results obtained by our methodology. In addition to location-based deviations, the generic shape aligns, demonstrating the reasonableness of our data.}
    \label{fig:dailyprofile_comp}
\end{figure}

A critical aspect of heating demand time series for planning applications is providing reasonable intra-day resolution. However, methods like the HDD approach do not offer this, and data from physical models can result in unrealistic load peaks when the profiles are simply aggregated. To validate the hourly heating demand data from our method, Figure~\ref{fig:dailyprofile_comp} compares the average daily profile with a measured profile from Ref.~\cite{Koene2023}. The typical double-peak structure in the heating demand, as also reported in Ref.~\cite{Clegg2019}, is clearly visible, with the highest heating demand occurring in the morning and evening when most people are at home and temperatures are still low or have dropped again. In addition to this generic structure, the shape of the profile depends on the location of the city and the corresponding outside temperature profile. While the upper subplot in Figure~\ref{fig:dailyprofile_comp} shows data from the Netherlands (with generally lower daytime temperatures), the lower subplot displays the demand profile of a city in Spain. Despite this, the comparison demonstrates that our method provides a reasonable intra-day resolution.

\subsection{Comparison with other methodologies}
\begin{figure}
    \centering
    \includegraphics[width=1\linewidth]{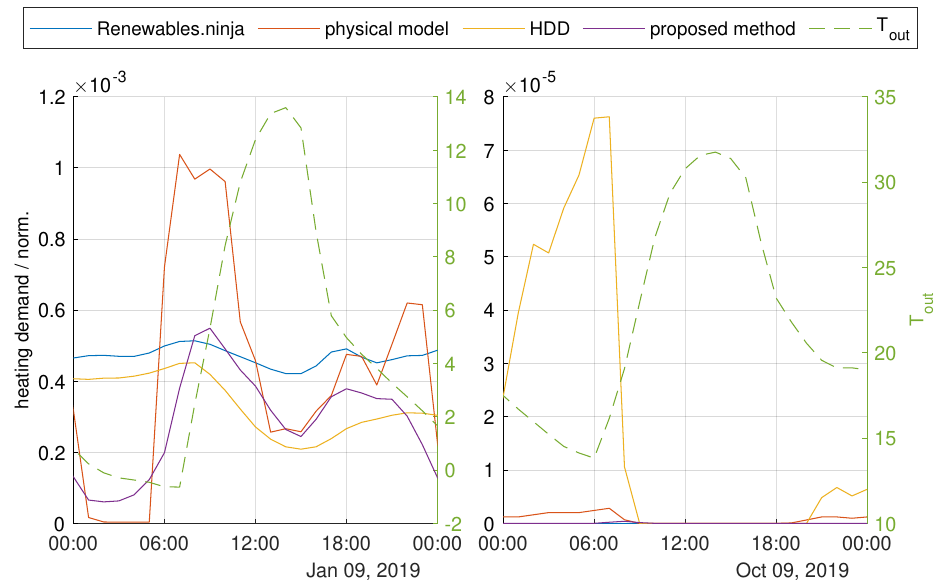}
    \caption{Comparison of different methodologies for generating heating demand profiles for a winter (left) and summer (right) day. Profiles are normalized to the yearly sum (same annual heating demand).}
    \label{fig:profile_comparison}
\end{figure}

As discussed in the Introduction~\ref{sc:Intro-tempmehtods}, a variety of different approaches for time series generation exists. In Figure~\ref{fig:profile_comparison}, we compare our results with data from a few commonly used methods, specifically:
\begin{itemize}
    \item HDD: standard heating degree days method, expanded for hourly resolution.
    \item Renewables.ninja: data from Ref.~\cite{Staffell}, calculated by an advanced heating degree days method, that includes additional parameters. 
    \item Physical model: data from a building simulation software. Simulations were performed for three building types, and added with weights, representing the share of the building type. 
\end{itemize}

The left tile shows demand during a winter day, where the shortcomings of simply aggregating data from physical models are evident. Due to the lack of a stochastic factor, all demand occurs simultaneously, producing unrealistic peaks. In the right tile, a late summer day is depicted: With warm temperatures during the day, all methods that consider thermal inertia result in no or very low heating demand. In contrast, the HDD method, which solely considers the outside temperature and neglects historical information, provides a non-zero value for the heating demand.

Both Renewables.ninja and our method address the systematic limitations mentioned above. However, it appears that Renewables.ninja underestimates the impact of user behavior, as the heating demand during nighttime is only slightly lower than during peak hours. In contrast, our model combines the strengths of both approaches by accurately accounting for thermal physics and the stochastic effects of diversity among different buildings. As a result, the data generated by our model seems to represent a convolution of the physical model and the outcomes from Renewables.ninja.

\subsection{Methodical and data uncertainty}
\label{sec:uncert}
Although our method produces accurate heating demand profiles at an aggregated level, it is important to emphasize that each independent load profile at the building level represents a random sample, derived from overall statistical data and assumed building properties. As such, it does not necessarily exactly match the heating demand of a specifically given building. Since the primary goal of this methodology is to generate data for large-scale analyses rather than to design individual heating systems for specific buildings, this limitation does not significantly affect the overall results.

As with any modeling approach, the quality of input data is critical to the accuracy of the output. For heating demand at the building level, uncertainties arise from incomplete or outdated data in GIS/OSM or cadaster systems, as well as from ambiguities in the assignment of building archetypes, as discussed by Dochev et al.~\cite{Dochev2020}. However, Dochev et al. also note that such discrepancies tend to average out when considering aggregated data. \sm{For a specific application in a case study, we strongly encourage researchers applying this method to carefully examine the input data and conduct plausibility checks and sample validations.}

Potential systematic errors may also arise from neglecting additional factors, such as solar irradiation, \sm{urban heat islands,} wind chill, internal heat gains from the human body, lighting, cooking, other electrical appliances, or varying individual setpoint temperatures and ventilation habits. Nonetheless, Staffell et al.~\cite{Staffell} demonstrate that, in their model, most of these factors have only a minor effect, with solar gain being the most significant. In comparison, the impact of these factors is generally smaller than the uncertainty introduced by the input data.

\sm{For projects that require precise solar gain calculations, more detailed building models are necessary. Accurately assessing shading effects and the amount of effective solar radiation reaching buildings requires consideration of building shapes, orientation, morphology, roof shapes and window openings \cite{Frayssinet2018}. Such analyses typically rely on highly detailed building data, including 3D models. These models are also crucial for evaluating the impact of urban heat islands \cite{Li2019a}, which, as reported by \cite{Yang2020}, can change the heating demand of individual buildings by up to 20\%. However, comprehensive 3D building data is not yet widely accessible to the public.} \sm{Therefore, we have deliberately chosen a simplified model, acknowledging a certain degree of uncertainty while ensuring broad and universal applicability of our approach. Beyond that,} it is worth noting that variations between different climate years may result in larger deviations than the factors discussed above.

\section{Case Study: Analyzing the heating demand of Puertollano/Spain}
\label{sc:casestudy}
% Descritpion of the situation: PV, Electroliser, etc. Citiy sizse & co

Following a brief motivation for selecting this particular case and an introduction to the context in Section~\ref{sc:casestudymotivation}, we describe the application of our methodology to Puertollano in Section~\ref{sc:casestudy_data}. The results obtained are presented and discussed in Section~\ref{sc:casestudy-results}, with an emphasis on the value of the datasets for smart planning applications. 

\sm{It is important to note that the applicability of the proposed method is not limited to this case study. This analysis illustrates the general structure of the obtained data and represents just one of many possible applications.}

\subsection{Motivation and aim of the case study}
\label{sc:casestudymotivation}
% general: find best way to decabonise heating in a district/city
% motivation why there, role of electrolyzers here and in EU in general (2x40GW). 
In 2022, one of the largest electrolyzers in Europe was commissioned in Puertollano, Spain. This electrolyzer, with a nominal electric input power of \SI{20}{\mega \watt}, is powered by a \SI{100}{\mega \watt} photovoltaic (PV) park located nearby~\cite{Puertollano_Iberola}. With the planned addition of \SI{200}{\mega \watt} of electrolyzer capacity, the region is set to become a key European hydrogen hub. As part of this project, it was proposed to utilize the waste heat from the electrolyzer, at a temperature of \SI{56}{\celsius}, for district heating to provide affordable heat to up to 3,000 households~\cite{newspaper_electrolyser_2022}.

Given the lower temperature level and the potential intermittency related to solar irradiation and electrolyzer operational schedule, a more advanced planning approach is required compared to traditional heat engineering methods for network design. Sector-coupled optimization models are a leading tool for assessing the technical and economic feasibility of such projects and determining the optimal system configuration. However, detailed heating demand time series at a granular spatial level are essential for comprehensive analysis.

In this case study, we demonstrate that our methodology is capable of generating such plausible synthetic load profiles at the building level, using publicly available data. Beyond the application for this particular analysis, these profiles serve as a reliable foundation for further analyses, such as, optimizing strategies for decarbonizing heating systems, or determining future additional electricity load profiles, particularly in scenarios involving the partial or full electrification of heating demand.

\subsection{Data analysis}
\label{sc:casestudy_data}
% how data was accessed (data sources) and how it was processed (Matalb, number of buildings, etc.). 
For the Puertollano case study, the initial dataset for the region was sourced from OSM~\cite{OSM}, which provides details on various infrastructures, including \num{11725} buildings, of these, 9,298 are residential buildings. Each building object contains information about the centroid location, shape (defined by GPS coordinates), and height recorded.

\begin{figure}
\centering
    \begin{subfigure}[b]{0.49\textwidth}
        \centering
        \includegraphics[width=0.99\textwidth]{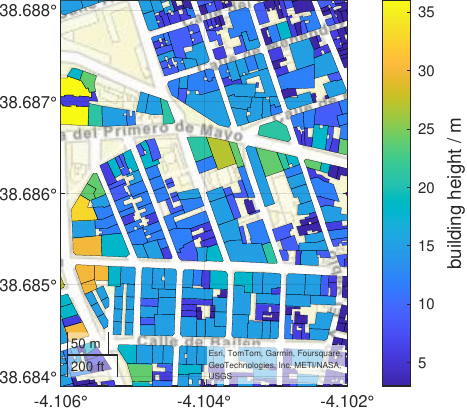}
        \caption{building height}
        \label{building_proerties_height}
    \end{subfigure}
    \begin{subfigure}[b]{0.49\textwidth}
        \centering
        \includegraphics[width=0.99\textwidth]{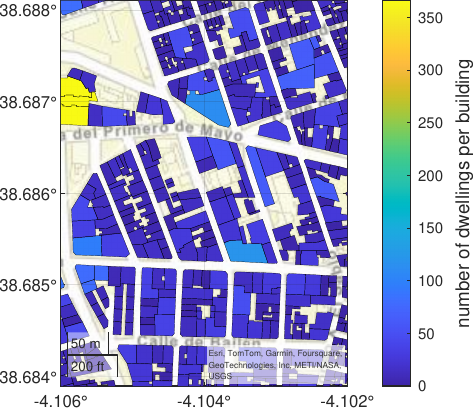}
        \caption{number of dwelling per building}
        \label{building_proerties_year}
    \end{subfigure}
    \begin{subfigure}[b]{0.49\textwidth}
        \centering
        \includegraphics[width=0.92\textwidth]{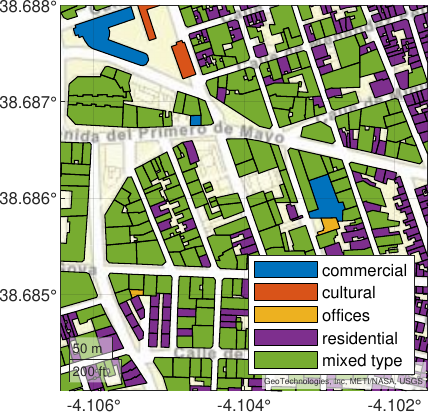}
        \caption{type of usage}
        \label{building_proerties_num}
    \end{subfigure}
    \begin{subfigure}[b]{0.49\textwidth}
        \centering
        \includegraphics[width=0.99\textwidth]{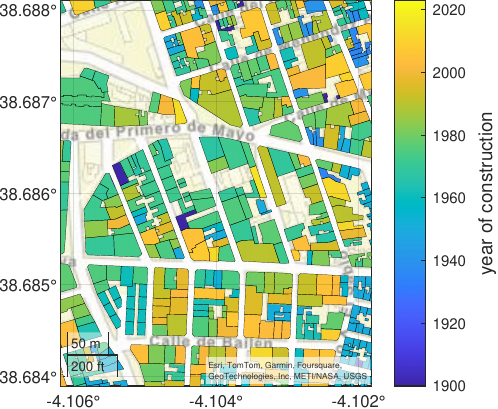}
        \caption{year of construction or last renovation}
        \label{building_proerties_type}
    \end{subfigure}
\caption{Different building properties of a small sector of the town, based on OSM and cadaster data.}
\label{fig:building_data}
\end{figure}

Subsequent to this, additional building-specific data was obtained from the Spanish cadaster~\cite{SedeEkectronicaCadastro} using the provided API\sm{~\cite{API_Cadaster}}. This dataset includes information on building usage type, construction year, area designation, number of units, and postal address. For buildings containing multiple units (e.g. apartment buildings), data for each individual subobject (household) was requested and aggregated at the building level. A sample of this data at the building level is presented in Figure~\ref{fig:building_data}. \sm{Due to the lack of detailed 3D building information and building-specific window areas, solar gains are not considered in this case study, such that $Q_\mathrm{solargain} = 0$ in Equation~\eqref{eq:Tin-iterative}. This simplification avoids introducing additional uncertainty from estimated or assumed solar exposure parameters, which would require complex geometric modeling and accurate material properties that are typically not available from open data sources.}

To estimate the specific annual heating demand, the information on the year of construction, number of units, and number of floors was combined with typological data from the TABULA Webtool~\cite{Loga2016}. For this analysis, a continental climate was selected, and heating demand data for both original and retrofitted buildings was extracted separately for further study.

In the next step, the thermal load profile for each building was computed. The thermal properties were estimated using the available data, as outlined in Section~\ref{sec:step4}. The MCMC method was then employed to generate an active occupancy profile, $A(t)$, with an hourly resolution for both business days and weekends. The transition matrices were derived from Ref.~\cite{Richardson2008}. For the outdoor temperature data, the dataset from Renewables.ninja~\cite{RENW_Ninja} was utilized. These inputs were subsequently processed through the thermal model, where temperatures and heating demands were iteratively calculated over 8760 steps. This procedure resulted in an hourly heating demand time series for each building. The same process was also applied using data for the nominal heating demand of the buildings, assuming they were retrofitted. The generated dataset can be accessed in Ref.~\cite{dataexample}. \sm{The total computational time for calculating the time series for the city of \num{11725} buildings and 8760 hours is approximately \SI{69}{\second} on a standard laptop (Intel i7, 32 GB RAM). A detailed overview of the computational performance can be found in the Supplementary Material~\cite{supply}.}

\subsection{Results}
\label{sc:casestudy-results}
% what are the results: what is the peak demand, the total yearly demand. 

\begin{figure}
    \centering
    \includegraphics[width=\textwidth]{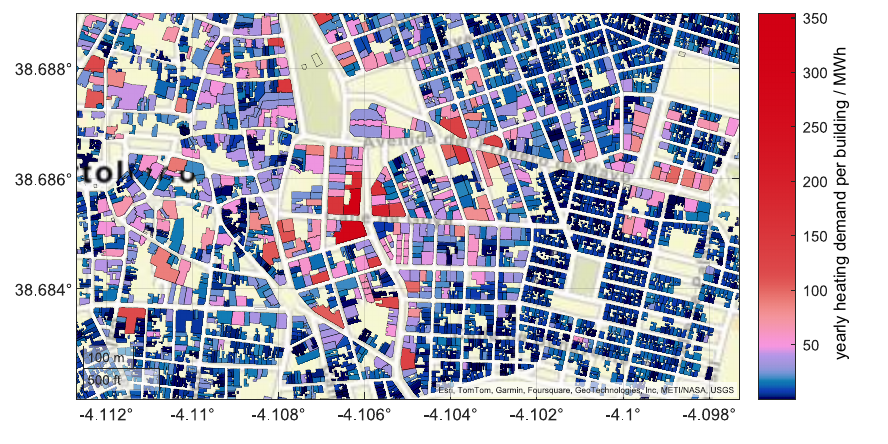}
    \caption{Calculated total annual heating demand per building with the proposed methodology.}
    \label{fig:tot-heatdem}
\end{figure}

Figure~\ref{fig:tot-heatdem} presents the calculated total annual heating demand at the building level. The total residential heating demand for the city amounts to \SI{114.2}{\giga \watt \hour} per year. This value aligns well with the results from the Spanish heat map~\cite{MapDeCalor} and the data from the Hotmaps project~\cite{Hotmaps}. By analyzing the building-level data, we can observe that \SI{41}{\percent} of the total heating demand is attributed to just 1020 buildings, which represent \SI{10}{\percent} of the total number of buildings.

\begin{figure}
    \centering
    \includegraphics[width=\textwidth]{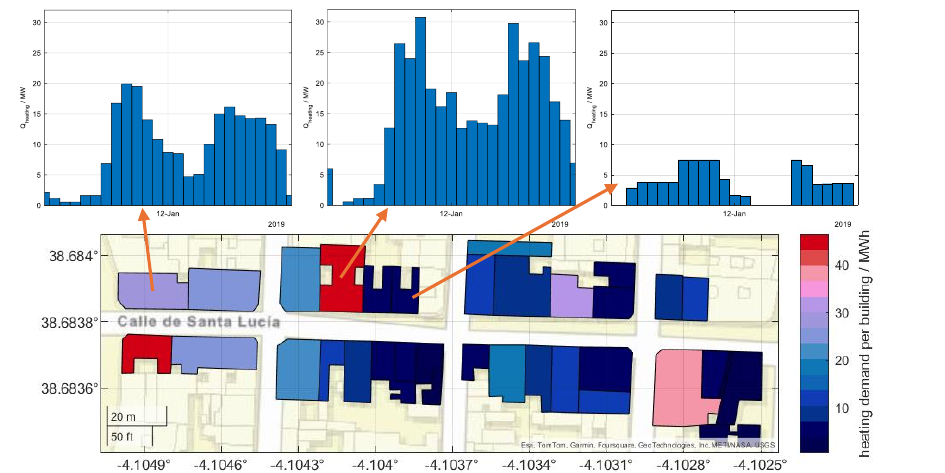}
    \caption{Illustration of considering diversity among different heating demand profiles and their convergence towards averaged profiles for an increasing number of buildings. The individual profiles depend on the thermal properties of the building and the assigned active occupancy profile.}
    \label{fig:profilessumup}
\end{figure}

The aggregation of individual heating demand profiles is shown in Figure~\ref{fig:profilessumup}. Each building has a unique temporal demand profile, reflecting the diverse daily routines of different societal groups. As more individual profiles are summed up, the resulting curve converges towards a smooth double-peak structure, consistent with patterns observed in the literature~\cite{Clegg2019,Koene2023}.

\begin{figure}
    \centering
    \includegraphics[width=\textwidth]{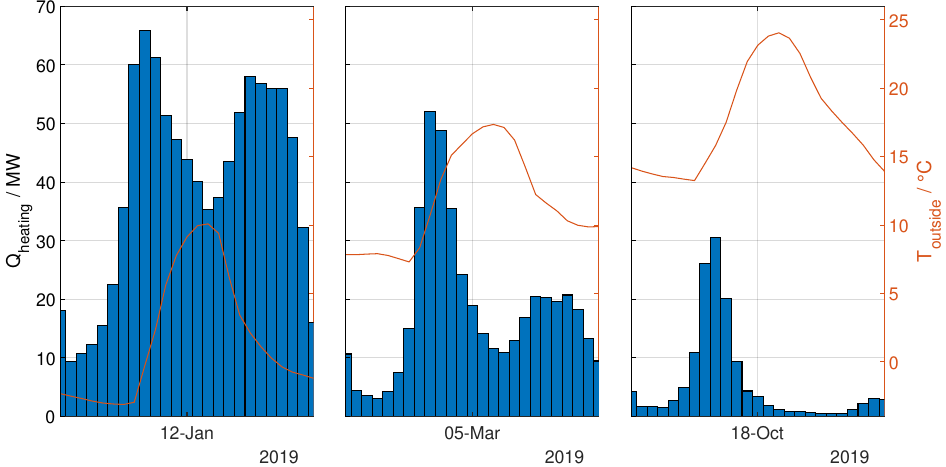}
    \caption{Aggregated heating demand time series for three different days (0 -- 24 hours) and corresponding outdoor temperature.}
    \label{fig:total-load-profiles}
\end{figure}

The total load profile for all buildings in Puertollano is presented in Figure~\ref{fig:total-load-profiles} for various times throughout the year. The peak heating demand reaches \SI{65.8}{\mega \watt} on a morning in January (the 273rd hour of the year), coinciding with the coldest day of the year.

To illustrate one advantage of using time series data at the building level, we compare two different thermal renovation strategies. First, we randomly select \SI{15}{\percent} of the buildings (weighted by residential area; so in total, \SI{15}{\percent} of the total residential area is renovated) for thermal retrofitting. For these buildings, improved thermal properties are applied (based on Ref.~\cite{Loga2016}), and the thermal heating demand time series are recalculated. This process is repeated 50 times with different randomly selected buildings to obtain a statistical distribution. The result is a new yearly demand of \SI{108.3 \pm 0.3}{\giga \watt \hour} (a reduction of \SI{5.2}{\percent}) and a peak demand of \SI{62.1 \pm 2}{\mega \watt} (a reduction of \SI{5.6}{\percent}).

We now repeat the analysis deliberately choosing buildings based on their initial thermal performance instead of randomly.
% For comparison, we repeat the analysis, but instead of selecting buildings randomly, we deliberately choose buildings based on their initial thermal performance.
By applying energy-saving measures to \SI{15}{\percent} of the buildings (weighted by residential area) with the worst thermal performance (according to the revised Energy Efficiency Directive EU/2023/1791~\cite{energy_eff_directive}), we achieve a reduction of \SI{13}{\percent} in yearly demand and a \SI{15}{\percent} reduction in peak demand.

\begin{figure}
    \centering
    \includegraphics[width=\textwidth]{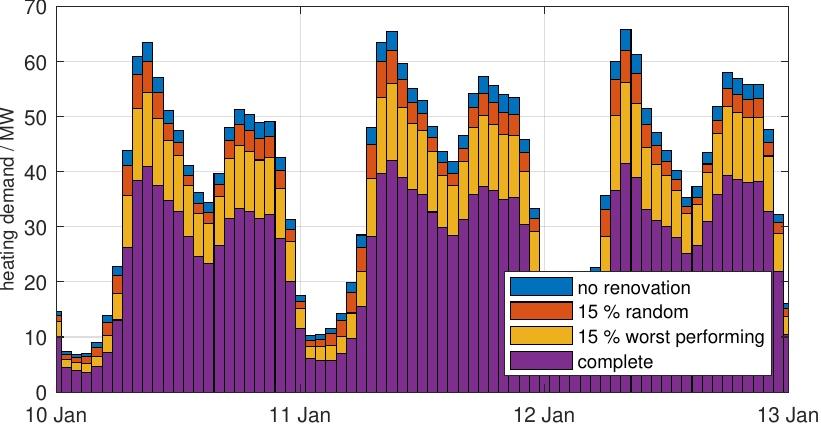}
    \caption{Comparison of influence of different renovation strategies on the thermal demand time series.}
    \label{fig:retrofitting}
\end{figure}

This suggests that smart and selective renovation strategies could be more than twice as effective as uncoordinated actions, which aligns with the findings of Ref.~\cite{PradesGil2023}. Figure~\ref{fig:retrofitting} compares the time series for the different renovation strategies. It can be seen that the smart approach also contributes to flattening the peak in thermal demand, potentially allowing for a smaller, more cost-effective district heating network design.

\section{Conclusion and Outlook}
\label{sc:conclusion}
% it was shown that solely based on the simple data very realistic output can be generated. 

In this study, we presented and validated an integrated methodology that generates heat load profiles at the building level, enabling accurate aggregation of individual building data. A key advantage of this approach is the minimal input data required, which allows for broad applicability across Europe. As a result, this method can be used to create datasets that provide highly resolved yearly heating demands and aggregated load profiles for districts on a granular scale.

Such data is invaluable for a variety of applications in smart energy planning. It can help identify clusters of buildings with poor thermal performance for targeted renovation initiatives, facilitate sector-coupled optimization models to decarbonize the residential sector, match intermittent waste heat potentials with optimal heating demand profiles, support district heating grid operators in more detailed planning for grid expansion or optimization, and assist electrical grid operators in better predicting the impact of ongoing electrification of the heating sector on electricity load profiles in the distribution grid. The method also offers full temporal resolution flexibility, assuming outdoor temperature and occupancy profiles are available at the desired time scale.

In the future, improved data availability, particularly regarding the thermal performance of buildings, will further increase the accuracy of our proposed methodology.
% Future improvements to the methodology will be enabled by better data availability, particularly regarding the thermal performance of buildings.
For example, many dwellings already have precise thermal performance calculations in energy performance certificates, and such data is sometimes available in public registers (e.g.,~\cite{EnergieAusweisDB_AT}). Making this information accessible for research projects could significantly enhance data quality, though appropriate measures to preserve data privacy must be implemented. Additionally, incorporating other factors such as solar irradiation, wind chill, or city microclimates could additionally improve the model’s accuracy.

\FloatBarrier

%TC:ignore
\section*{CRediT authorship contribution statement}
\textbf{Simon Malacek:} Conceptualization, Investigation, Methodology, Validation, Visualization, Writing - original draft. \textbf{José Portela:} Project administration, Supervision, Validation. \textbf{Yannick Werner} Supervision, Writing - review and editing. \textbf{Sonja Wogrin:} Supervision, Resources, Writing - review and editing. 
%TC:endignore

\section*{Data Availability}
\label{sc:dataavailability}

The dataset generated for the case study \sm{and for the validation study} can be accessed in Ref.~\cite{dataexample}.
     
%TC:ignore
\section*{Acknowledgements}
Many thanks to David Cardona Vasquez and Alexander Michael Konrad for their preliminary work in this field and the insightful discussions. 

This research was supported by funding from CDTI, with Grant Number MIG-20221006 associated with the ATMOSPHERE Project.

\section*{Declaration of generative AI and AI-assisted technologies in the writing process} 
During the preparation of this work, the authors used ChatGPT from OpenAI and Grammarly from Grammarly Inc. to enhance readability and perform grammar and spell-checking. After using these tools, the authors reviewed and edited the content. The authors take full responsibility for the content of this publication.

\bibliographystyle{elsarticle-num} 
\bibliography{simons_libary}
%\bibliography{simons_libary}



\end{document}